\documentclass{ws-mpla}

\usepackage[super]{cite}
\usepackage{xcolor}
\usepackage[verbose,hypertexnames=false]{hyperref}
\hypersetup{colorlinks=true,allbordercolors=blue,pdfborderstyle={/S/U/W 1}}
\begin{document}

\markboth{Al-Harthi and Stevenson}{Surface Energy in Nuclear Octupole Excitations}

%%%%%%%%%%%%%%%%%%%%% Publisher's Area please ignore %%%%%%%%%%%%%%
\catchline{}{}{}{}{}
%%%%%%%%%%%%%%%%%%%%%%%%%%%%%%%%%%%%%%%%%%%%%%%%%%%%%%%%%%%%%%%%%%%

\title{The role of the surface energy in nuclear octupole excitations}

\author{Khlood Alharthi}

\address{Department of Physics, College of Science and Humanities, Jubail, Imam Abdulrahman Bin Faisal University, Jubail, 35811, Saudi Arabia\\
  and\\
  University of Surrey, Guildford, Surrey, GU2 7XH, United Kingdom\\
k.alharthi@surrey.ac.uk}

\author{Paul Stevenson}

\address{University of Surrey, Guildford, Surrey, GU2 7XH, United Kingdom\\
p.stevenson@surrey.ac.uk}
\maketitle

\begin{history}
\received{(Day Month Year)}
\revised{(Day Month Year)}
\accepted{(Day Month Year)}
\published{(Day Month Year)}
\end{history}

\begin{abstract}
  Octupole excitations of atomic nuclei can be viewed as fluctuations around an equilibrium shape.  Shape fluctuations cause a change in the nuclear surface, and it is reasonable to suggest that such fluctuations may probe the nuclear surface energy.  We use a series of Skyrme interactions, which were fitted to provide a systematic range of surface energies, to explore the surface energy dependence of octupole excitations in $^{208}$Pb.  We find a strong positive linear correlation between the surface energy of a Skyrme interaction and its prediction of the first $3^-$ octupole excitation energy.
\end{abstract}


%\ccode{PACS Nos.: 03.65.$-$w, 04.62.+v}

\section{Introduction}
The representation of bulk properties of infinite or finite nuclear matter in terms of coefficients in mass formulae remains a useful way of classifying nuclei and the forces that act in them.  A basic form of the semi-empirical mass formula, following Bohr and Mottelson's textbook \cite{bohrmottelson}, Weizs\"acker \cite{weiz}, and Bethe and Bacher \cite{bethebacher} can be expressed as

\begin{equation}
\mathcal{B} = a_{\mathrm{vol}}A - a_{\mathrm{surf}}A^{2/3} -\frac{1}{2}a_{\mathrm{sym}}\frac{(N-Z)^2}{A}-\frac{3}{5}\frac{Z^2e^2}{R_c}\pm\delta(N,Z),\label{eq:semf}
\end{equation}
where $\mathcal{B}$ is the total binding energy of a nucleus of $N$ neutrons and $Z$ protons, with total mass number $A=N+Z$.  The coefficient $a_{\mathrm{vol}}$ characterises the volume term, accounting for the nearest-neighbour interaction of each nucleon with its surrounding nucleons.  The $a_\mathrm{surf}$ surface coefficient corrects the volume term for those nucleons which lack a full set of nearest-neighbour due to being at the surface of the nucleus.  The $a_\mathrm{sym}$ term accounts for the preference for a like number of protons and neutrons in the nucleus while the final term, in which $e$ is the elementary charge and $R_c$ the charge radius of the nucleus, gives a contribution from the Coulomb force.  Finally, the $\pm\delta(N,Z)$ gives a correction due to the pairing interaction, differentiating odd and even proton and neutron numbers.

From a fit to observed binding energies, numerical values of each coefficient can be obtained.  For example, the value for $a_{\mathrm{surf}}$ quoted by Bohr and Mottelson is $\simeq 17$ MeV, and a recent fit to a modern set of mass data gives $a_s=16.329509 \pm 0.146924$ \cite{ned}.  One may then go on to link this macroscopic view with any microscopic theory that is able to calculate these coefficients and other derived properties alongside more detailed structure, thereby linking structure details to the bulk properties \cite{stone,colo18}.  

The surface energy of finite nuclear matter has been studied relatively less than other terms, as it does not apply in the calculationally simple case of infinite nuclear matter.  However, it has recently become of increasing interest as the ability to make useful calculations of e.g. heavy-ion reactions \cite{hias} and fission\cite{pancic} have become possible with effective interactions in the density functional framework.  This has combined with the availability of parameters sets for Skyrme effective interactions in which the surface energy is systematically varied \cite{jodon,islam}, while other terms (e.g. volume) are held constant.

The purpose of the present work is to study the effect of the systematic variation of the surface energy on shape vibrations in the form of the lowest octupole state in doubly-magic $^{208}$Pb.  The rationale is that the vibration can be thought of as dynamical shape-changing processes in which the ratio of surface-to-volume of the nucleus changes, as one expects as a spherical object (dynamically) undergoes deformation~\cite{oss}.  The idea is that by using a series of nuclear effective interactions in which the surface energy varies while the volume term is fixed, the vibrational states will be sensitive to the variation in the surface energy.  Other effects captured in the semi-empirical mass formula e.g. Coulomb term should be minimally significant across the chain of interactions.  The Coulomb force itself is fixed outside of the nuclear interaction, and will depend solely on the proton shape and its dynamic vibration, which is fixed by the choice of octupole vibration. 

The octupole state in $^{208}$Pb was selected over any other particular choice as an example system to study here since $^{208}$Pb is a doubly-magic nucleus, spherical in its ground state, with a well-defined and well-studied first excited state~\cite{kib02} which can be well-characterised as a shape vibration.  Furthermore, as a heavy nucleus with an internal region of constant density \cite{negele}, $^{208}$Pb has a definite ``volume'' element, as well as a surface.  Arguably, light nuclei are all surface. 

\section{Methodology}
\label{sec:method}

We employ the time-dependent density functional theory, through the published Sky3d code \cite{sky3d10,sky3d12} using the set of Skyrme interactions developed by Jodon {\it et al.}\cite{jodon}.  The time-dependent density functional theory (TDDFT) \cite{simenel} is the general microscopic theory of quantum dynamics truncated at the one-body level.  Formally, theories such as the random phase approximation (RPA) can be derived from it \cite{simenel}, and the practical correspondence between TDDFT and RPA for the case of resonance states has been demonstrated \cite{burrello}.  %While RPA is a more widespread theory used in practice for microscopic calculation of vibrational excitations, we use TDDFT to get a real-time picture of the shape changes when a nucleus is undergoing a vibrational excitation.

The set of Skyrme forces used in our study were fitted so that each had identically-constrained values of many infinite nuclear matter and finite nuclear properties, while being adjusted to particular values of the surface energy -- essentially the $a_{\mathrm surf}$ coefficient of equation \ref{eq:semf}.  Since extracting the $a_{\mathrm surf}$ parameter from a Skyrme interaction depends on the particular method used, Jodon {\it et al.} presented different values for the surface energy, depending on method.  For our analysis we equate our $a_{\mathrm surf}$ with their $a_{\mathrm{surf}^{\mathrm{(HF)}}}$, extracted using Hartree-Fock calculations of a slab of semi-infinite nuclear matter.

In order to calculate the octupole vibrational state in $^{208}$Pb, we first run the Sky3d code in static Hartree-Fock mode to generate a ground state, then apply an instantaneous octupole boost to each single particle wave function $\psi(r,t)$ of the form

\begin{equation}
\psi(r,0^+) = \exp^{ikr^3Y_{30}}\psi(r,0), \label{eq:boost}
\end{equation}
where $\psi(r,0)$ are the static ground state wave functions.  $r^3Y_{30}$, where $Y_{30}$ is a spherical harmonic, is the octupole excitation operator and initialises the octupole shape excitation.   Once the boost is applied, the single-particle states are evolved forward in time by the TDDFT equations, and the nucleus allowed to respond to the octupole excitation.  The octupole operator $O(t)$ is followed in time to analyse the response

\begin{equation}
  O(t) = \langle r^3Y_{30}\rangle(t).
\end{equation}

The nuclear strength is given by

\begin{equation}
  S(E) = \frac{1}{\hbar k}\tilde{O}(\omega) \label{eq:strength},
\end{equation}
where $\tilde{O}(\omega)$ is the Fourier Transform of the time-dependent response and $k$ is the boost strength parameter featured in equation (\ref{eq:boost}).  In this work, we stick to the linear RPA limit with a boost strength of $k=0.0001$ fm$^{-3}$.  Note that the exact value of $k$ is not important as long as one is in the linear regime.  The strength, eq (\ref{eq:strength}), divides the response signal by $k$.  In the linear regime, the response scales as $k$.  For further details see the review by Simenel \cite{simenel}.

In order to extract peak energies from a finite time signal, the data is enveloped as described in ref. \cite{sky3d12}, with a linear interpolation made between the discretised energy grid points to estimate the peak location and height.

\section{Results}\label{sec:results}

For each of the eight SLy5s$X$ forces with $X=1\ldots8$, we calculated the response of the $^{208}$Pb nucleus to an octupole boost, and performed the Fourier analysis.  The tabulated peak positions for each force, along with the given surface energy, is shown in Table \ref{tab:results}.  Note that the values of $a_{\mathrm{surf}}$ are those calculated by Jodon et al. in their fits \cite{jodon}. 

\begin{table}[ht]
\tbl{Lowest octupole state in $^{208}$Pb for each SLy5s$X$ force\label{tab:results}}
{\tabcolsep13pt\begin{tabular}{@{}ccc@{}}
\toprule
Force Name & Surface Energy $a_{\mathrm{surf}}^{\mathrm{(HF)}}$ [MeV] & $E_{3^-}$ [MeV] \\
\colrule
SLy5s1 & 17.55 & 3.122\\
SLy5s2 & 17.74 & 3.153\\
SLy5s3 & 17.93 & 3.182\\
SLy5s4 & 18.12 & 3.211\\
SLy5s5 & 18.31 & 3.241\\
SLy5s6 & 18.50 & 3.268\\
SLy5s7 & 18.70 & 3.298\\
SLy5s8 & 18.89 & 3.323\\
\botrule
\end{tabular}}
\end{table}

\begin{figure}[ht]
\centerline{\includegraphics[width=0.9\textwidth]{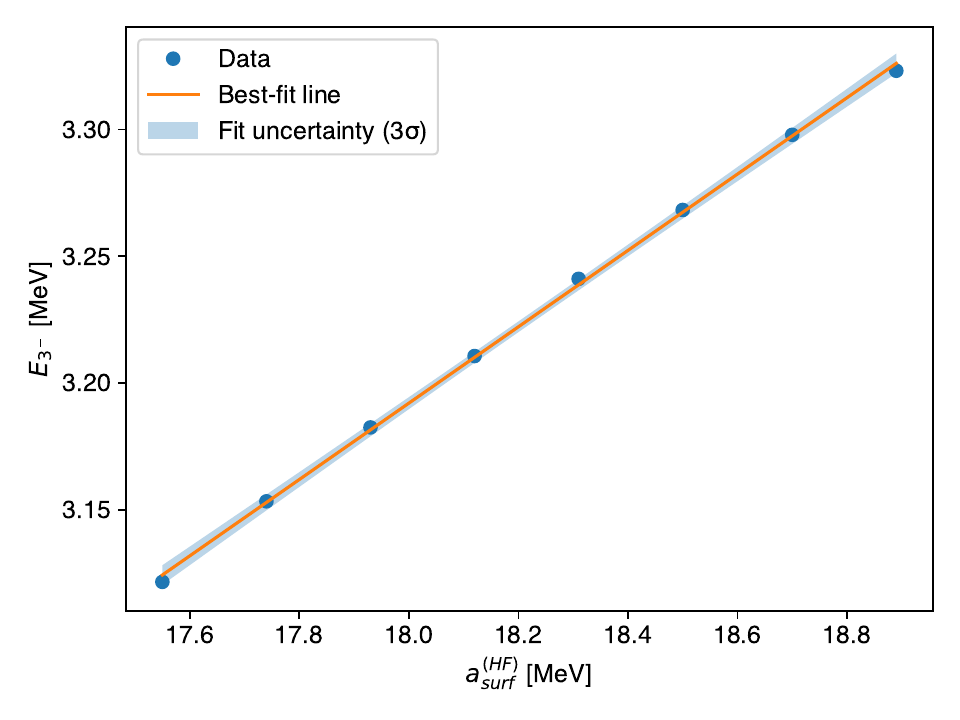}}
\vspace*{8pt}
\caption{Correlation between the surface energy and the predicted octupole peak position for each Skyrme interaction in the SLy5s$X$ series.}
\label{fig:fit}
\alttext{straight line fit of the surface energy to the octupole peak}
\end{figure}

The correlation between the surface energy and the $3^-$ energy is shown in Figure \ref{fig:fit}.  Here, the straight line fit is shown.  It is made with linear regression with the resulting equation

\begin{equation}
  E_{3^-} = 0.1506 a_{\mathrm{surf}}^{(HF)} + 0.4812\ {\mathrm{MeV}}. \label{eq:fit}
\end{equation}

The plot includes a $3\sigma$ envelope from the fit.  The strong linearity in the fit means that the error band is small, and the high uncertainty envelope of $3\sigma$ was chosen to render it visible on the scale of the plot.  

\begin{figure}[ht]
\centerline{\includegraphics[width=0.9\textwidth]{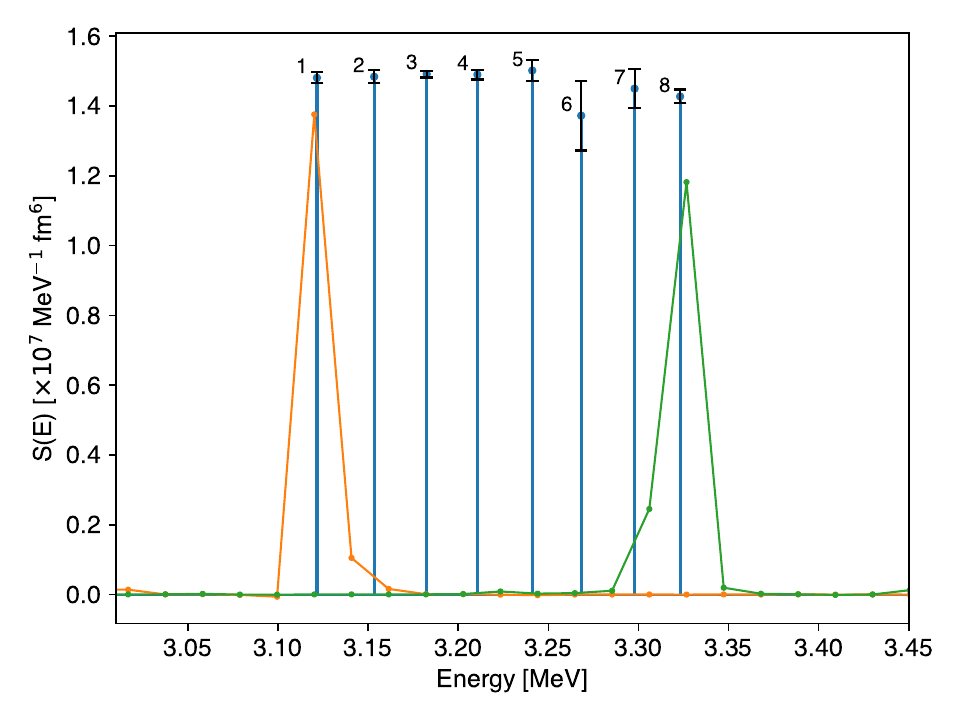}}
\vspace*{8pt}
\caption{Octupole strength in $^{208}$Pb for the forces SLy5s$X$ forces for $X=1..8$ in the region of the low-lying excited 3$^-$ state.  Peak positions and strengths are shown for all eight forces.}
\label{fig1}
\alttext{A plot of theoretical results for the octupole state in lead-208 using a set of Skyrme forces with varying surface properties}
\end{figure}

To further visualise the results, Figure \ref{fig1} shows the strength function as described by equation (\ref{eq:strength}).  Here, the peaks are shown at the points calculated through the interpolation method discussed in section \ref{sec:method}.  The peak height includes an error bar given by the height of the highest neighbouring point in the Fourier transform not included in the interpolation.  For the forces SLy5s1 and SLy5s8 the raw Fourier transformed strength is shown which shows both the discretisation of the energy grid caused by the finite sampling time of the octupole signal, and the spreading of strength between neighbouring energy points caused by the envelope technique \cite{sky3d12} used in the Fourier analysis. 

\section{Discussion and Conclusion}

The results from section \ref{sec:results} suggest that the there is a strong linear relationship between surface energy and the position of the octupole state in $^{208}$Pb.  We hypothesise that this is due to the fact that an octupole vibration involves a change in the surface area of a nucleus, and where the cost of generating the nuclear surface is low because of a low surface energy coefficient, the corresponding excitation energy of the nuclear state is lower.

The experimental value for the first $3^-$ energy in $^{208}$Pb is 2614 keV \cite{henderson}.  This is substantially lower than the values coming from any of the SLy5s$X$ forces.  We do not suggest here that an extrapolation of surface energy for this particular level makes sense, as other systematic effects in the Skyrme fits should be taken into account, with existing Skyrme forces showing a wide variation in reproduction of this energy level \cite{taqi}, as well as the existence of strong evidence for the importance of the details of the single-particle levels \cite{skx,twitter} in reproducing spectra.  However, the linear fit shown in equation (\ref{eq:fit}) appears robust, and could form a low-cost way of including pseudodata in future fits or studies.  However, alongside the recent fit of $a_\mathrm{surf}\simeq16.33$ MeV cited above, it does suggest that an extension of the well-controlled fits of Skyrme interactions to lower values of $a_\mathrm{surf}$ would be useful. 

We suppose that the case of strong correlation between the surface energy and excitation energy of the $3^-$ state in $^{208}$Pb extends to other nuclei and other shape vibrations.  A rich seam of octupole vibrational states exists in heavy nuclei, though many are not spherical, and this may decrease the effect \cite{butler}.  Other states, such as quadrupole vibrations, present throughout the nuclear chart in most spherical even-even nuclei \cite{verney}, could also be a fertile ground for study to further the picture explored in the present work.

In conclusion, we hypothesised a correlation between the surface energy which characterises an effective nuclear interaction, and the position of shape-vibrational excited octupole states.  We show there to be a strong linear correlation between the two quantities in the case of the well-known first excited state in $^{208}$Pb, which is octupole in nature. 

%\appendix
%\section{An Appendix Section}
\section*{Data Availability}
All data produced in this work was produced with published codes and using parameters as noted in the manuscript.  No further data than that presented is required for reproduction. 

\section*{Acknowledgments}
We acknowledge useful discussions with Jirina Stone.  KA-H gratefully acknowledges the financial support provided by Imam Abdulrahman Bin Faisal University. The support is sincerely appreciated.  PS gratefully acknowledges financial support from the UK STFC funding agency under grant ST/Y000358/1.  For the purpose of open access, the author has applied a Creative Commons attribution license (CC BY) to any Author Accepted Manuscript version arising from this submission.  

\section*{ORCID}

\noindent Khlood Al-Harthi - \url{https://orcid.org/0000-0003-3063-9131}\\
\noindent Paul Stevenson - \url{https://orcid.org/0000-0003-2645-2569}

%\section*{References}

\end{document}